# MORPHOLOGICAL INSTABILITY AND CANCER INVASION:
## A 'SPLASHING WATER DROP' ANALOGY

[1] Caterina Guiot, [2] Pier Paolo Delsanto and [3,*] Thomas S. Deisboeck

[1] Dip. Neuroscience and CNISM, Università di Torino, Italy.
[2] Dip. Fisica, Politecnico di Torino, Italy.
[3] Complex Biosystems Modeling Laboratory, Harvard-MIT (HST) Athinoula A. Martinos Center for Biomedical Imaging, Massachusetts General Hospital, Charlestown, MA 02129, USA.

**\*Corresponding Author:**

Thomas S. Deisboeck, M.D.
Complex Biosystems Modeling Laboratory
Harvard-MIT (HST) Athinoula A. Martinos Center for Biomedical Imaging
Massachusetts General Hospital-East, 2301
Bldg. 149, 13th Street
Charlestown, MA 02129
Tel: 617-724-1845
Fax: 617-726-7422
Email: deisboec@helix.mgh.harvard.edu

**Abstract**

We present an analogy between two unrelated instabilities. One is caused by the impact of a drop of water on a solid surface while the other concerns a tumor that develops invasive cellular branches into the surrounding host tissue. In spite of the apparent abstractness of the idea, it yields a very practical result, i.e. an index that predicts tumor invasion based on a few measurable parameters. We discuss its application in the context of experimental data and suggest potential clinical implications.

**Background**

Tissue invasion is one of the hallmarks of cancer [1]. From the primary tumor mass, cells are able to move out and infiltrate adjacent tissues by means of degrading enzymes (e.g., [2]). Depending on the cancer type, these cells may form distant settlements, i.e. metastases (e.g., [3]). Tumor expansion therefore results from the complex interplay between the developmental ability of the tumor itself and the characteristics of the host tissue in which its growth occurs (e.g., [4]).

**Figure 1**

It has been recently proposed [5] that cancer invasion can be described as a morphological *instability* that occurs during solid tumor growth and results in invasive 'fingering', i.e. branching patterns (see **Figure 1**). This instability may be driven by any physical or chemical condition (oxygen, glucose, acid and drug concentration gradients), provided that the average cohesion among tumor cells decreases and/or their adhesion to the stroma increases (for a recent review on related molecular aspects, such as the cadherin-'switch', see [6]). In fact, the aforementioned model of Cristini et al. [5] shows that reductions in the *surface tension* at the tumor-tissue interface may generate and control tumor branching in the nearby tissues. A previous investigation from the same group [7] had analyzed different tumor growth regimes and shown that invasive fingering *in vivo* could be driven by vascular and elastic anisotropies in highly vascularized tumors. A recent advance [8] shows that the competition between proliferation (shape-destabilizing force) and adhesion (shape-stabilizing force) can be implemented in a more general mathematical description of tumor growth and accounts for many experimental evidences. Correspondingly, another recent paper by Anderson [9] stresses the relevance of cell adhesion in the process of tumor invasion.

Undoubtedly, a more detailed insight into the mechanisms that drive tumor invasion is critical for targeting these cancer cell populations more effectively, and, possibly, concepts derived from other scientific disciplines may contribute valuable insights. It is in this line of thought that we propose an *analogy* with the case of a liquid drop, which impacts on a solid surface and causes the formation of a fluid 'crown'. Such instability, termed Rayleigh or Yarin-Weiss capillary instability, has been extensively studied in the field of fluid dynamics [10-12] (see **Figure 2**).

**Figure 2**

Both phenomena share similar features, such as secondary jets (corresponding to the invasive branching in the case of tumors), nucleation near the fluid rim (corresponding to the evidence for branching confluence) and dispersion of small drops at the fluid-air interface (with resemblance to proliferating aggregates that have been reported to emerge within the invasive cell population [13].

**The Model**

Fluid dynamicists describe their system by means of some non-dimensional numbers, such as the Weber number We $= \rho\, D\, V^2 / \sigma$ , the Ohnesorge number Oh $= \mu / \text{sqrt}(\rho\, \sigma\, D)$, the Reynold number Re $= \rho\, D\, V / \mu$  and the Capillary number Ca$= \mu\, V / \sigma$ , where $\rho$ is the fluid density, D the drop diameter, V the impact velocity, $\mu$  the fluid viscosity and $\sigma$ the surface tension. For instance, [14] showed that the splash/non-splash boundary for several different fluids is well described by sqrt(Ca) = 0.35. Provided an estimate for both tumor viscosity and surface tension is available, it would be interesting to investigate whether similar non-dimensional quantities could discriminate between invasive and non invasive tumor behaviour. Moreover, also the number of invasive branches may be predicted on the basis of the previous nondimensional numbers [15]. According to these authors the number of branches is given by:

$$N_f = 2\, \pi\, R / \lambda \tag{1}$$
$$\lambda = 2\, \pi\, (3\, \sigma / \rho\, a\, )^{1/2} \tag{2}$$

where R is the radius of the drop and *a* is  the 'deceleration' at the impact.  In the case of a splashing drop, the impact velocity is the main measurable parameter related to the way in which kinetic energy is converted into surface energy, associated with the increased free-surface area and viscous dissipation. Splashing is in some sense the droplet's reaction to a sudden increase in pressure. The relevant quantity, due to the very short time scale involved in the process, is the splashing impact.

In the case of an invading tumor, however, the concept of 'impact' cannot be used; the increased mechanical pressure, exerted by the confining microenvironment due to cancer expansion, elicits a much slower response, hence the instability develops over a much larger time scale. This suggests that, equivalently, the tumor cell-matrix interaction could be a critical parameter. We therefore propose that the deceleration $a$ can be evaluated starting from the confining mechanical pressure P exerted by the host tissue on the growing tumor. Assuming for simplicity a spherical shape for the tumor, where S and V are the surface and volume at the onset of invasion, respectively, we obtain:

$$a = P\,S/\,\rho\,V = 3\,P\,/\,\rho\,R \qquad\qquad (3)$$

It follows that

$$N_f = (PR/\,\sigma)^{\;\frac12} \qquad\qquad (4)$$

The value $N_f = 1$ separates the case of no branching (hence no tumor invasion) from that in which at least one branch develops (and invasion takes place). By defining the dimensionless *Invasion Parameter*, IP, as

$$IP = PR\,/\sigma, \qquad\qquad (5)$$

we deduce that, provided IP <1 (which implies large surface tension, small confining pressure and/or radius value) tumor invasion cannot occur, while for IP>1 invasive behaviour is expected (see **Figure 3**).

**Figure 3**

This index defines a critical value that predicts different potential outcomes of tumor growth instability regimen, i.e. self-similar growth versus invasive branching. The evaluation of the actual extent and/or rate of invasion may involve many other parameters which characterize the tumor growth processes as well as the microenvironment. In particular, provided the extent of invasion is related to the number of invasive branching, Eqn. (4) relates the 'invasion efficiency' to the square root of IP.

## Discussion

We appreciate the obvious differences between tumor biology and fluid dynamics. Yet, solely on the basis of the aforementioned perceived *analogy*, we argue here that the morphological instability that drives tumor invasion is controlled by a dimensionless parameter (IP) which is proportional to the confining pressure and tumor radius, yet inversely proportional to its surface tension. As a consequence, increasing levels of confinement at larger tumor radii should *promote* the onset of invasion, while larger values of adhesion-mediated surface tension can inhibit it. The former is in agreement with our previous, experimentally driven notion of a feedback between the key characteristics of proliferation and invasion [16] and the argument for a quantitative link between them [17]. The model's parameters can be measured and monitored, as well as modified with a treatment regimen. Intriguingly, the following ongoing experimental investigations seem already to confirm our conjectures:

**a)** *Tumor surface tension:* Winters et al. [18] have investigated three different cells lines derived from malignant astrocytoma (U-87MG, LN-229 and U-118MG). Their work shows that (i) surface tension in the multicell aggregates they have used is independent of the compressive forces and that the spheres practically behave as a liquid and not as elastic aggregates; (ii) the measured aggregate surface tension is about 7 dyne/cm for U-87Mg, 10 for LN-229 and more than 16 for U-118MG, and (iii) there is indeed a significant *inverse* correlation between invasiveness and surface tension; (iv) finally, the anti-invasive therapeutic agent Dexamethasone increases the microscopic tumor's surface tension or cohesivity amongst cells in direct contact. For the cell lines studied above, surface tension is therefore a predictor for *in vitro* invasiveness and the authors suggest a threshold value for σ of about 10 dyne/cm. We add that other aspects of surface tension and intercellular adhesion have also been investigated, and shown to be relevant for non invasive tumor development (e.g. [19-21]).

**b)** *Microenvironmental pressure:* Some papers have recently addressed the problem of the mechanical interaction of the matrix with the embedded tumor. For instance, Paszek et al. [22] claim that stiffer tissues are expected to promote malignant behaviour. Also, an experimental investigation by Georges and Janmey [23] shows that a basic NIH3T3 fibroblast embedded in a soft polyacrylamide gel develops with a roughly spherical shape (suggesting prevalence of cohesive forces), while in a stiff gel it exhibits finger-like features (consistent with the preponderance of adhesive forces). Another recent paper, by Kaufman et al. [24], investigates how glioblastoma spheroids grow in and invade 3D collagen I matrices, differing in collagen concentration and thus, in their average stiffness (the elastic modulus of the 0.5 mg/ml gel was 4 Pa, in the 1.0 mg/ml gel was 11 Pa and in the 1.5 mg/ml was about 100 Pa). Using *in vitro* microscopy techniques, the authors show that in the 0.5 mg/ml gel there are few invasive cells around the tumor. At larger concentrations (1.5-2.0 mg/ml gels) invasion occurs more quickly and the number of invasive cells increases. In conclusion, reducing the matrix stiffness seems to reduce the number of invasive cells and their invasion rate. It is noteworthy in this context that the relevance of local pressure in hindering the growth of non-invasive MTS has been studied by e.g. [25-27].

**c)** *Tumor radius*: Tamaki et al. [13] investigated C6 astrocytoma spheroids with different diameters (i.e., 370, 535 and 855 µm on average) that were implanted in collagen type I gels. The authors showed that spheroid size indeed correlated with a larger total invasion distance and an increased rate of invasion. We note that, reflecting the complexity of the cancer system's expansion process properly [16, 17], our concept relies on experimental conditions that allow for *both* cancer growth *and* invasion to occur. From Eqn. (5) it follows that, if invasion is restricted by the chosen experimental conditions and $\sigma$ is assumed to remain constant, any increase in P beyond a certain threshold would result in limiting R. This is indeed confirmed by Helmlinger et al. [25], who reported that a solid stress of 45-120 mmHg inhibits the growth of multicellular tumor spheroids cultured in an agarose matrix (according to the authors, 'cells cannot digest or migrate through it').

## Conclusions

In summary, our model, while admittedly very simple, suggests - based on a striking fluid dynamics analogy - several clinical management strategies that, separately or in combination, should yield anti-invasive effects. They include (aside from the obvious initial attempt to *reduce the tumor size* through surgical techniques and accompanying non-surgical approaches (radio- and chemotherapy)):

**(1)** *Promoting tumor cell-tumor cell adhesion and thus increasing the tumor surface tension* $\sigma$. Interestingly, experiments on prostate cancer cells have already shown that stable transfection of E-cadherin (the prototype cell-cell adhesion molecule that is increasingly lost with tumor progression) results in cellular cohesiveness and a decrease in invasiveness, in part due to a down-regulation of matrix metalloproteinase (MMP) activity [28][1].

**(2)** *Reducing the confining mechanical pressure exerted on the tumor.* This refers to pharmacological strategies that range from applying perioperatively corticosteroids, as it is standard for treating malignant brain tumors [30], to preventing pressure-stimulated cell adhesion, i.e. mechanotransduction by targeting the cytoskeleton's actin polymerization [31, 32].

Taken together, our model is not only supported by a variety of experimental findings, but it offers already an explanation for the anti-invasive and anti-metastatic effects seen in the aforementioned experimental studies and clinical regimen, respectively. As such, this model has the potential to further our understanding of the dynamical relationship between a tumor and its microenvironment, and, in its future iterations, may even hold promise for assessing the potential impact of combinatory treatment approaches.

**Competing Interests:** The authors declare that they have no competing interests.

**Authors Contributions:** All authors contributed equally to this work. All have read and approved the final manuscript.

**Acknowledgements**: This work has been supported in part by NIH grant CA 113004 and by the Harvard-MIT (HST) Athinoula A. Martinos Center for Biomedical Imaging and the Department of Radiology at Massachusetts General Hospital.

---

[1] Such a functional relationship (and thus our argument to capitalize on it for therapeutic purposes) is further supported by results from squamous cell carcinoma cells that had been genetically engineered to stably express a dominant-negative E-cadherin fusion protein [29]. The authors reported that, in three-dimensional environments, E-cadherin deficiency indeed led to a loss of intercellular adhesion and triggered tumor cell invasion by MMP-2 and MMP-9 driven matrix degradation.

## References

1. Hanahan D, Weinberg RA: **The hallmarks of cancer**. *Cell* 2000, **100**: 57-70.

2. Nakano A, Tani E, Miyazaki K, Yamamoto Y, Furuyama J: **Matrix metalloproteinases and tissue inhibitors of metalloproteinases in human gliomas**. *J. Neurosurg* 1995, **83**: 298-307.

3. Fidler IJ: **The pathogenesis of cancer metastasis: the 'seed and soil' hypothesis revisited**. *Nat Rev Cancer* 2003, **3**: 453–458.

4. Postovit L-M, Seftor EA, Seftor REB, Hendrix MJC: **Influence of the microenvironment on melanoma cell fate determination and phenotype**. *Cancer Res* 2006, **66**: 7833-7836.

5. Cristini V, Frieboes HB, Gatenby R, Caserta S, Ferrari M, Sinek J: **Morphologic instability and cancer invasion**. *Clin Cancer Res* 2005, **11**: 6772-6779

6. Christofori G: **New signals from the invasive front**. *Nature* 2006, **441**: 444-450.

7. Cristini V, Lowengrub J, Nie Q: **Nonlinear simulation of tumor growth**. *J. Math. Biol*. 2003, **46**:191-224.

8. Frieboes HB, Zheng X, Sun C, Tromberg B, Gatenby R, Cristini V: **An integrated computational/experimental model of tumor invasion.** *Cancer Res* 2006, **66**:1597-1604.

9. Anderson ARA: **A hybrid mathematical model of solid tumour invasion: the importance of cell adhesion**. *Math Med and Biol* 2005, **22**: 163-186.

10. Rozhkov A, Prunet-Foch B, Vignes-Adler M: **Dynamics of a liquid lamella resulting from the impact of a water drop on a small target**. *Proc R Soc Lond A* 2004, **460**: 2681-2704.

11. Rioboo R, Bauthier C, Conti J, Voue M, De Coninck J: **Experimental investigation of splash and crown formation during single drop impact on wetted surfaces**. *Exp in Fluids* 2003, **35**: 648-652.

12. Yarin AL: **Drop impact dynamics: Splashing, spreading, receding, bouncing.** *Ann Rev Fluid Mech* 2006, **38**: 159-192.

13. Tamaki M, McDonald W, Amberger VR, Moore E, Del Maestro RF: **Implantation of C6 astrocytoma spheroid into collagen type I gels: invasive, proliferative, and enzymatic characterizations**. *J. Neurosurg* 1997, **87**: 602-609.

14. Vander Wal RL, Berger GM, Mozes SD: **The combined influence of a rough surface and thin fluid film upon the splashing threshold and splash dynamics of a droplet impacting onto them**. *Exp in Fluids* 2006, **40**: 53-59.

15. Bussmann M, Chandra S, Mostaghimi J: **Modeling the splash of a droplet impacting a solid surface**. *Phys Fluids* 2000, **12**: 3123-3132.

16. Deisboeck TS, Berens ME, Kansal AR, Torquato S, Stemmer-Rachamimov AO, Chiocca EA: **Pattern of self-organization in tumour systems: complex growth dynamics in a novel brain tumour spheroid model**. *Cell Prolif* 2001, **34**: 115-134.

17. Deisboeck TS, Mansury Y, Guiot C, Degiorgis PG, Delsanto PP: **Insights from a novel tumor model: Indications for a quantitative link between tumor growth and invasion**. *Med. Hypotheses* 2005, **65**: 785-790.

18. Winters BS, Shepard SR, Foty RA: **Biophysical measurements of brain tumor cohesion**. *Int J Cancer* 2005, **114**: 371-379.

19. Greenspan HP: **On the growth and stability of cell cultures and solid tumours**. *J. theor. Biol.* 1976, **56**: 229-242.

20. Byrne HM:. **The importance of intercellular adhesion in the development of carcinomas**. *IMA J. Math. Appl. Med. Biol.* 1997, **14**: 305-323

21. Chaplain MAJ, Sleeman BD: **Modelling the growth of solid tumours and incorporating a method for their classification using nonlinear elasticity theory**. *J. Math. Biol.* 1993, **31**: 431-473.

22. Paszek MJ, Zahir N, Johnson KR, Lakins JN, Rozenberg GI, Gefen A, Reinhart-King CA, Margulies SS, Dembo M, Boettiger D, Hammer DA, Weaver VM: **Tensional homeostasis and the malignant phenotype**. *Cancer Cell* 2005, **8**: 241-254.

23. Georges PC, Janmey PA: **Cell type-specific response to growth on soft materials**. *J Appl Physiol* 2005, **98**: 1547-1553.

24. Kaufman LJ, Brangwynne CP, Kasza KE, Filippidi E, Gordon VD, Deisboeck TS, Weitz DA: **Glioma expansion in collagen I matrices: analyzing collagen concentration-dependent growth and motility patterns**. *Biophysical J* 2005, **99**: 636-650.

25. Helmlinger G, Netti PA, Lichtenbeld HC, Melder RJ, Jain RK: **Solid stress inhibits the growth of multicellular tumor spheroids**. *Nature Biotech* 1997, **15**: 778-783.

26. Chen CY, Byrne HM, J.R. King JR: **The influence of growth-induced stress from the surrounding medium on the development of multicell spheroid**. *J. Math. Biol.* 2001, **43**: 191-220.

27. Roose T, Netti PA, Munn LL, Boucher Y, Jain RK: **Solid stress generated by spheroid growth estimated using a linear poroelasticity model**. *Microvasc. Res.* 2003, **66**: 204-212.

28. Luo J, Lubaroff DM, Hendrix MJC: **Suppression of prostate cancer invasive potential and matrix metalloproteinase activity by E-cadherin transfection**. *Cancer Res* 1999, **59**: 3552-3665.

29. Margulis A, Zhang W, Alt-Holland A, Crawford HC, Fusenig NE, Garlick JA: **E-cadherin suppression accelerates squamous cell carcinoma progression in three-dimensional, human tissue constructs**. *Cancer Res* 2005, **65**: 1783-1791.

30. Chang SM, Parney IF, Huang W, Anderson FA Jr, Asher AL, Bernstein M, Lillehei KO, Brem H, Berger MS, Laws ER: **Patterns of care for adults with newly diagnosed malignant glioma.** *JAMA* 2005, **293**: 557-564.

31. Thamilselvan V, Basson MD: **Pressure activates colon cancer cell adhesion by insight-out focal adhesion complex and actin cytoskeleton signalling**. *Gastroenterology* 2004, **126**: 8-18.

32. Thamilselvan V, Basson MD: **The role of the cytoskeleton in differentially regulating pressure-mediated effects on malignant colonocyte focal adhesion signalling and cell adhesion**. *Carcinogenesis* 2005, **26**: 1687-1679.

33. Habib S, Molina-Paris C, Deisboeck TS: **Complex dynamics of tumors: modeling an emerging brain tumor system with coupled reaction-diffusion equations**. *Physica A* 2003, **327**: 501-524.

## Figure Legends

**Figure 1.** Microscopy image of a multicellular tumor spheroid, exhibiting an extensive branching system that rapidly expands into the surrounding extracellular matrix gel. These branches consist of multiple invasive cells. (From Habib et al. [33], with permission).

**Figure 2.** Water drop impact on a solid surface**.** (Courtesy Adam Hart-Davis/DHD Multimedia Gallery at http://gallery.hd.org/index.jsp)

**Figure 3.** The surface IP = 1 according to Eqn. (5).

**Figures**

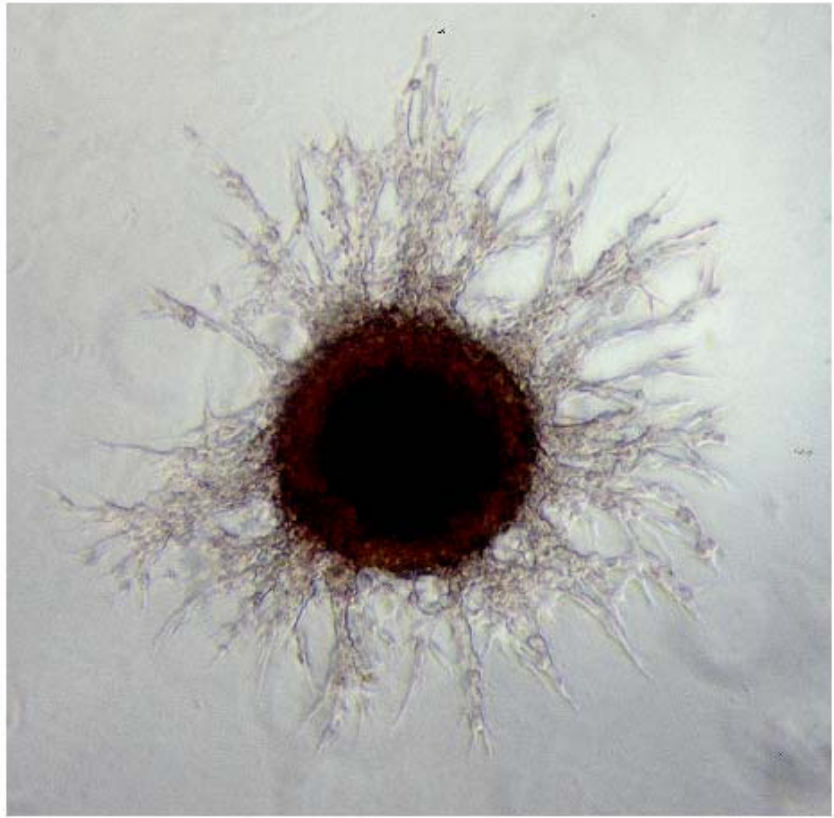

**Figure 1.**

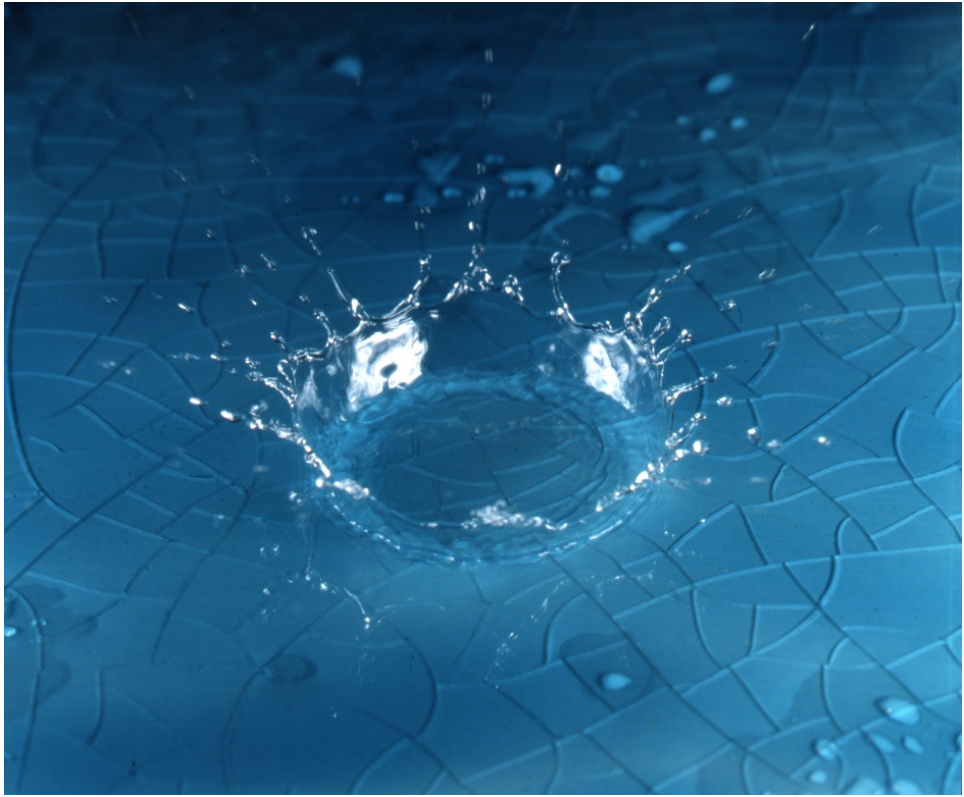

**Figure 2.**

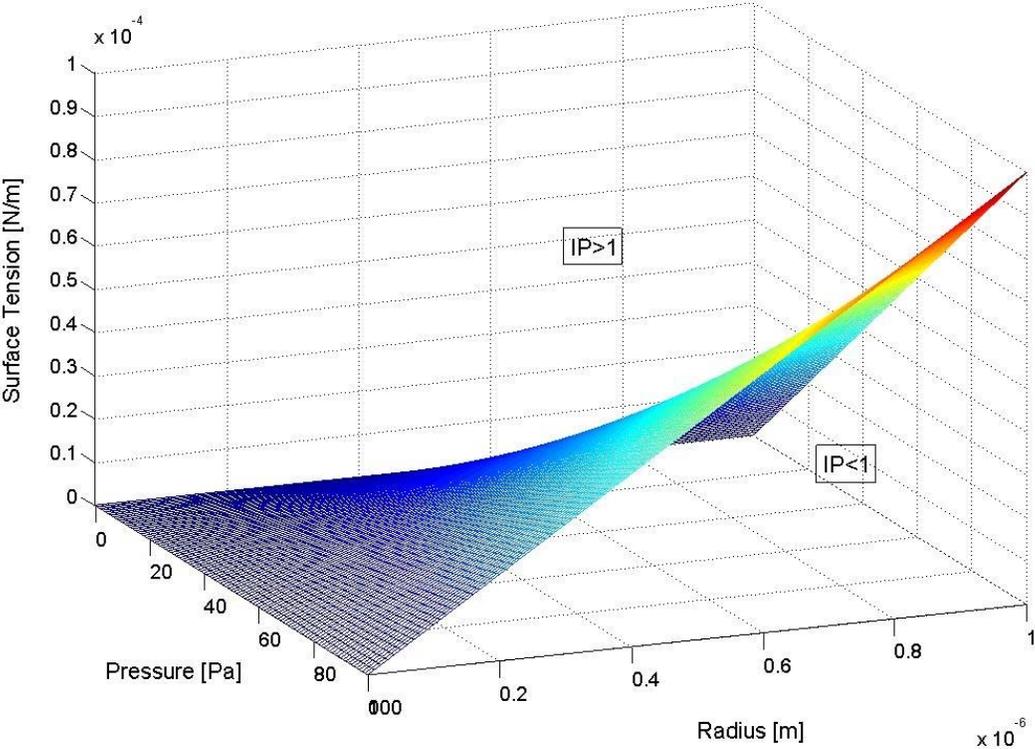

**Figure 3.**